\documentclass[12pt]{article}
\usepackage{epsf}
\title{ {\bf
$\tau\rightarrow \mu \, \bar{\nu_i} \, \nu_i$ decay in the general
two Higgs doublet model}}

\author{\vspace{1cm}\\
        {\bf E. O. Iltan}
        \thanks{E-mail address:
        eiltan@heraklit.physics.metu.edu.tr}\,\,\, and
   {\bf H. Sundu}
        \thanks{E-mail address:
        sundu@.metu.edu.tr}
 \\
        Physics Department, Middle East Technical University \\
        Ankara, Turkey\\}
\date{}

\begin{document}
\setlength{\baselineskip}{24pt}
\maketitle
\setlength{\baselineskip}{7mm}
\begin{abstract}
We study $\tau\rightarrow \mu \,\bar{\nu_i}\, \nu_i$,
$i=e,\mu,\tau$ decay in the model III version of the two Higgs
doublet model. We calculated the $BR$ at the order of the
magnitude of $10^{-6}-10^{-4}$ for the intermediate values of the
Yukawa couplings. Furthermore, we predict the upper limit of the
coupling for the $\tau-h^0 (A^0)-\tau$ transition as $\sim 0.3$ in
the case that the $BR$ is $\sim 10^{-6}$. We observe that the
experimental result of the process under consideration can give
comprehensive information about the physics beyond the standard
model and the free parameters existing.
\end{abstract}
\thispagestyle{empty}
\newpage
\setcounter{page}{1}
\section{Introduction}
Lepton flavor violating (LFV) interactions are interesting since
they do not exist in the standard model (SM) and give strong
signal about the new physics beyond. Such decays have reached
great interest at present and the experimental search has been
improved. Among LFV decays the ones existing in the leptonic
sector are clean theoretically in the sense that they are free
from the nonperturbative effects. The processes $\mu\rightarrow
e\gamma$, $\tau\rightarrow e (\mu) \gamma$, $\tau\rightarrow e
\bar{e} e$, $\tau\rightarrow e \bar{\mu} \mu$, are the examples of
LFV interactions. There are on-going and planned experiments for
$\mu\rightarrow e\gamma$ ($\tau\rightarrow \mu \gamma$) and the
current limits for their branching ratios ($BR$) are $1.2\times
10^{-11}$ \cite{Brooks} ($1.1\times 10^{-6}$ \cite{Ahmed}). The
numerical estimates predict that the $BR$ of the proceses
$\tau\rightarrow e \bar{e} e$, $\tau\rightarrow e \bar{\mu} \mu$
are at the order of the magnitude of $10^{-6}$ \cite{Ilakovac},
which is a measurable value in the LEP experiments and $\tau$
factories.

In such decays, the assumption of the non-existence of
Cabibbo-Kobayashi-Maskawa (CKM) type matrix in the leptonic sector
forbids the charged Flavor Changing (FC) interactions and
therefore, the physics beyond the SM plays the main role, where
the general two Higgs doublet model (2HDM), so called  model III,
is one of the candidate. This model predicts that the  LFV
interactions can exist at loop level and the internal neutral
higgs bosons $h_0$ and $A_0$ play the main role. There are number
of Yukawa couplings which describe the strength of the
interactions lepton-lepton-neutral Higgs particle, appearing in
the loops and their strength can be determined by the experimental
data. In the literature, there are several studies on LFV
interactions in different models. Such interactions are studied in
a model independent way in \cite{Chang}, in the framework of model
III 2HDM \cite{ilt}, in supersymmetric models \cite{Barbieri1,
Barbieri2,Barbieri3,Ciafaloni,Duong,Couture,Okada}.

\begin{figure}[htb]
\vskip 0.0truein \centering \epsfxsize=6.8in
\leavevmode\epsffile{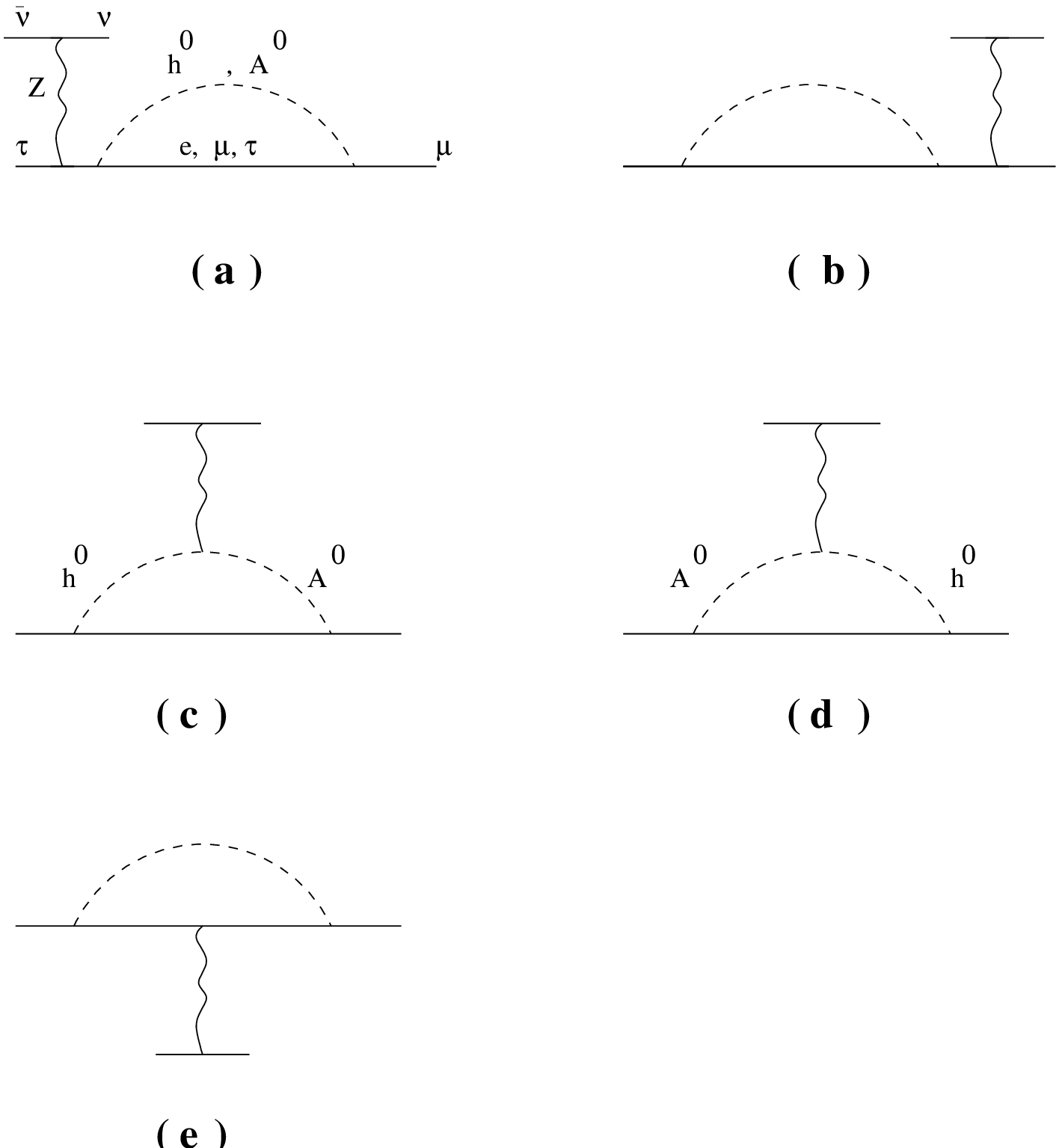} \vskip 1.0truein \caption[]{One loop
diagrams contribute to $\tau\rightarrow \mu \,\bar{\nu_i}\,
\nu_i$, $i=e,\mu,\tau$ decay due to the neutral Higgs bosons $h_0$
and $A_0$ in the model III version of 2HDM. Solid lines represent
leptons and neutrinos, curly (dashed) lines represent the virtual
$Z$ boson ($h_0$ and $A_0$ fields).} \label{fig1}
\end{figure}

Our work is devoted to the study of the $\tau\rightarrow \mu
\bar{\nu_i} \nu_i$, $i=e,\mu,\tau$ decay in the model III version
of 2HDM. This process exists with the help of internal scalar
bosons $h^0$ and $A^0$ to obtain the flavor changing transition
$\tau\rightarrow\mu$ and the internal $Z$ boson to get the output
$\bar{\nu}\nu$ (see Fig. \ref{fig1}). The $BR(\tau\rightarrow \mu
\,\bar{\nu_i}\, \nu_i)$, $(i=e,\mu,\tau)$ is predicted at the
order of the magnitude of $10^{-6}-10^{-4}$ for the intermediate
values of the Yukawa couplings, which are the free parameters of
the model used and is strongly sensitive to the couplings for
$\tau-h^0 (A^0)-\tau$ and $\tau-h^0 (A^0)-\mu$ transitions. This
can play an important role in the determination of the upper
limits of these couplings, especially the one for the $\tau-h^0
(A^0)-\tau$ transition. Notice that there have been some
experimental studies on this process in the literature
\cite{experim}.

The paper is organized as follows: In Section 2, we present the
theoretical expression for the decay width of the LFV decay
$\tau\rightarrow \mu \,\bar{\nu_i}\, \nu_i$, $i=e,\mu,\tau$, in
the framework of the model III. Section 3 is devoted to discussion
and our conclusions.
\section{$\tau\rightarrow \mu \, \bar{\nu_i} \, \nu_i$ decay in the general
two Higgs doublet model}
Type III 2HDM permits flavor changing neutral currents (FCNC) at
tree level. The  Yukawa interaction for the leptonic sector in the
model III is
\begin{eqnarray}
{\cal{L}}_{Y}=
\eta^{E}_{ij} \bar{l}_{i L} \phi_{1} E_{j R}+
\xi^{E}_{ij} \bar{l}_{i L} \phi_{2} E_{j R} + h.c. \,\,\, ,
\label{lagrangian}
\end{eqnarray}
where $i,j$ are family indices of leptons, $L$ and $R$ denote chiral
projections $L(R)=1/2(1\mp \gamma_5)$, $\phi_{i}$ for $i=1,2$, are the
two scalar doublets, $l_{i L}$ and $E_{j R}$ are lepton doublets and
singlets respectively.
Here $\phi_{1}$ and $\phi_{2}$ are chosen as
\begin{eqnarray}
\phi_{1}=\frac{1}{\sqrt{2}}\left[\left(\begin{array}{c c}
0\\v+H^{0}\end{array}\right)\; + \left(\begin{array}{c c}
\sqrt{2} \chi^{+}\\ i \chi^{0}\end{array}\right) \right]\, ;
\phi_{2}=\frac{1}{\sqrt{2}}\left(\begin{array}{c c}
\sqrt{2} H^{+}\\ H_1+i H_2 \end{array}\right) \,\, ,
\label{choice}
\end{eqnarray}
with the vacuum expectation values
\begin{eqnarray}
<\phi_{1}>=\frac{1}{\sqrt{2}}\left(\begin{array}{c c}
0\\v\end{array}\right) \,  \, ;
<\phi_{2}>=0 \,\, .
\label{choice2}
\end{eqnarray}
By considering the gauge and $CP$ invariant Higgs potential which
spontaneously breaks  $SU(2)\times U(1)$ down to $U(1)$  as:
\begin{eqnarray}
V(\phi_1, \phi_2 )&=&c_1 (\phi_1^+ \phi_1-v^2/2)^2+ c_2 (\phi_2^+
\phi_2)^2 \nonumber \\ &+& + c_3 [(\phi_1^+ \phi_1-v^2/2)+
\phi_2^+ \phi_2]^2 + c_4 [(\phi_1^+ \phi_1)
(\phi_2^+ \phi_2)-(\phi_1^+ \phi_2)(\phi_2^+ \phi_1)] \nonumber \\
&+& c_5 [Re(\phi_1^+ \phi_2)]^2 + c_{6} [Im(\phi_1^+ \phi_2)]^2
+c_{7} \, , \label{potential}
\end{eqnarray}
with constants $c_i, \, i=1,...,7$, $H_1$ and $H_2$ are obtained
as the mass eigenstates $h^0$ and $A^0$ respectively, since no
mixing occurs between two CP-even neutral bosons $H^0$ and $h^0$
in the tree level. Therefore it is possible to collect the SM
particles in the first doublet and the new particles in the second
one. The part which produce FCNC at tree level is
\begin{eqnarray}
{\cal{L}}_{Y,FC}=  \xi^{E}_{ij} \bar{l}_{i L} \phi_{2} E_{j R} +
h.c. \,\, . \label{lagrangianFC}
\end{eqnarray}
Here the Yukawa matrices $\xi^{E}_{ij}$ have  complex entries in
general. Notice that,in the following, we replace $\xi^{E}$ with
$\xi^{E}_{N}$ where "N" denotes the word "neutral".

Now, we consider the lepton flavor changing process
$\tau\rightarrow \mu \,\bar{\nu} \, \nu$ and we expect that the
main contribution to this decay comes from the neutral Higgs
bosons, namely, $h_0$ and $A_0$ in the loop level, in the leptonic
sector of the model III, (see Fig. \ref{fig1}). The general
effective vertex for the interaction of off-shell Z-boson with a
fermionic current is obtained as
\begin{equation}
\Gamma^{(REN)}_\mu (\tau\rightarrow \mu Z^*)=  f_1 \, \gamma_\mu +
f_2  \,\gamma_\mu \gamma_5+f_3\, \sigma_{\mu\nu}
k^\nu+f_4\,\sigma_{\mu\nu}\gamma_5 k^\nu  \label{GammaRen2}
\end{equation}
where $k$ is the momentum transfer, $k^2=(p-p')^2$, $p$
($p^{\prime}$) is the four momentum vector of incoming (outgoing)
lepton. Taking into account all the masses of internal ($m_i$) and
external leptons ($m_{l_1}, \,m_{l_2}$), the explicit expressions
for the functions $f_1$, $f_2$, $f_3$ and $f_4$ are
\begin{eqnarray}
f_1&=& \frac{g}{64\,\pi^2\,cos\,\theta_W} \int_0^1\, dx \,
\frac{1}{m^2_{l_2}-m^2_{l_1}} \Bigg \{ c_V \, (m_{l_2}+m_{l_1})
\nonumber \\
&\Bigg(& (-m_i \, \eta^+_i + m_{l_1} (-1+x)\, \eta_i^V)\, ln \,
\frac{L^{self}_ {1,\,h^0}}{\mu^2}+ (m_i \, \eta^+_i - m_{l_2}
(-1+x)\, \eta_i^V)\, ln \, \frac{L^{self}_{2,\, h^0}}{\mu^2}
\nonumber \\ &+& (m_i \, \eta^+_i + m_{l_1} (-1+x)\, \eta_i^V)\,
ln \, \frac{L^{self}_{1,\, A^0}}{\mu^2} - (m_i \, \eta^+_i +
m_{l_2} (-1+x) \,\eta_i^V)\, ln \, \frac{L^{self}_{2,\,
A^0}}{\mu^2} \Bigg) \nonumber \\ &+&
c_A \, (m_{l_2}-m_{l_1}) \nonumber \\
&\Bigg ( & (-m_i \, \eta^-_i + m_{l_1} (-1+x)\, \eta_i^A)\, ln \,
\frac{L^{self}_{1,\, h^0}}{\mu^2} + (m_i \, \eta^-_i + m_{l_2}
(-1+x)\, \eta_i^A)\, ln \, \frac{L^{self}_{2,\, h^0}}{\mu^2}
\nonumber \\ &+& (m_i \, \eta^-_i + m_{l_1} (-1+x)\, \eta_i^A)\,
ln \, \frac{L^{self}_{1,\, A^0}}{\mu^2} + (-m_i \, \eta^-_i +
m_{l_2} (-1+x)\, \eta_i^A)\, ln \, \frac{L^{self}_{2,\,
A^0}}{\mu^2} \Bigg) \Bigg \} \nonumber \\ &-&
\frac{g}{64\,\pi^2\,cos\,\theta_W} \int_0^1\,dx\, \int_0^{1-x} \,
dy \, \Bigg \{ m_i^2 \,(c_A\,
\eta_i^A-c_V\,\eta_i^V)\,(\frac{1}{L^{ver}_{A^0}}+
\frac{1}{L^{ver}_{h^0}}) \nonumber \\ &-& (1-x-y)\,m_i\, \Bigg(
c_A\,  (m_{l_2}-m_{l_1})\, \eta_i^- \,(\frac{1}{L^{ver}_{h^0}} -
\frac{1}{L^{ver}_{A^0}})+ c_V\, (m_{l_2}+m_{l_1})\, \eta_i^+ \,
(\frac{1}{L^{ver}_{h^0}} + \frac{1}{L^{ver}_{A^0}}) \Bigg)
\nonumber \\ &-& (c_A\, \eta_i^A+c_V\,\eta_i^V) \Bigg (
-2+(k^2\,x\,y+m_{l_1}\,m_{l_2}\, (-1+x+y)^2)\,
(\frac{1}{L^{ver}_{h^0}} +
\frac{1}{L^{ver}_{A^0}})-ln\,\frac{L^{ver}_{h^0}}{\mu^2}\,
\frac{L^{ver}_{A^0}}{\mu^2} \Bigg ) \nonumber \\ &-&
(m_{l_2}+m_{l_1})\, (1-x-y)\, \Bigg ( \frac{\eta_i^A\,(x\,m_{l_1}
+y\,m_{l_2})+m_i\,\eta_i^-}
{2\,L^{ver}_{A^0\,h^0}}+\frac{\eta_i^A\,(x\,m_{l_1} +y\,m_{l_2})-
m_i\,\eta_i^-}{2\,L^{ver}_{h^0\,A^0}} \Bigg )  \nonumber \\ &+&
\frac{1}{2} \,\eta_i^A\,ln\,\frac{L^{ver}_{A^0\,h^0}}{\mu^2}\,ln\,
\frac{L^{ver}_{h^0\,A^0}}{\mu^2}
\Bigg \}\,, \nonumber \\
f_2&=& \frac{g}{64\,\pi^2\,cos\,\theta_W} \int_0^1\, dx \,
\frac{1}{m^2_{l_2}-m^2_{l_1}} \Bigg \{ c_V \, (m_{l_2}-m_{l_1})
\nonumber \\
&\Bigg(& (m_i \, \eta^-_i + m_{l_1} (-1+x)\, \eta_i^A)\, ln \,
\frac{L^{self}_{1,\,A^0}}{\mu^2} + (-m_i \, \eta^-_i + m_{l_2}
(-1+x)\, \eta_i^A)\, ln \, \frac{L^{self}_ {2,\,A^0}}{\mu^2}
\nonumber \\ &+& (-m_i \, \eta^-_i + m_{l_1} (-1+x)\, \eta_i^A)\,
ln \, \frac{L^{self}_{1,\, h^0}}{\mu^2}+ (m_i \, \eta^-_i +
m_{l_2} (-1+x)\, \eta_i^A)\, ln \,
\frac{L^{self}_{2,\,h^0}}{\mu^2} \Bigg) \nonumber \\ &+&
c_A \, (m_{l_2}+m_{l_1}) \nonumber \\
&\Bigg(& (m_i \, \eta^+_i + m_{l_1} (-1+x)\, \eta_i^V)\, ln \,
\frac{L^{self}_{1,\, A^0}}{\mu^2}- (m_i \, \eta^+_i + m_{l_2}
(-1+x)\, \eta_i^V)\, ln \, \frac{L^{self}_{2,\,A^0}}{\mu^2}
\nonumber \\ &+& (-m_i \, \eta^+_i + m_{l_1} (-1+x)\, \eta_i^V)\,
ln \, \frac{L^{self}_{1,\, h^0}}{\mu^2} + (m_i \, \eta^+_i -
m_{l_2} (-1+x)\, \eta_i^V)\, \frac{ln \,
L^{self}_{2,\,h^0}}{\mu^2} \Bigg) \Bigg \} \nonumber \\ &+&
\frac{g}{64\,\pi^2\,cos\,\theta_W} \int_0^1\,dx\, \int_0^{1-x} \,
dy \, \Bigg \{ m_i^2 \,(c_V\,
\eta_i^A-c_A\,\eta_i^V)\,(\frac{1}{L^{ver}_{A^0}}+
\frac{1}{L^{ver}_{h^0}}) \nonumber \\ &-& m_i\, (1-x-y)\, \Bigg(
c_V\, (m_{l_2}-m_{l_1}) \,\eta_i^- + c_A\, (m_{l_2}+m_{l_1})\,
\eta_i^+ \Bigg) \,(\frac{1} {L^{ver}_{h^0}} -
\frac{1}{L^{ver}_{A^0}}) \nonumber \\ &+& (c_V\,
\eta_i^A+c_A\,\eta_i^V) \Bigg(-2+(k^2\,x\,y-m_{l_1}\,m_{l_2}\,
(-1+x+y)^2) (\frac{1}{L^{ver}_{h^0}}+\frac{1}{L^{ver}_{A^0}})-
ln\,\frac{L^{ver}_{h^0}}{\mu^2}\,\frac{L^{ver}_{A^0}}{\mu^2}
\Bigg) \nonumber \\ &-& (m_{l_2}-m_{l_1})\, (1-x-y)\, \Bigg(
\frac{\eta_i^V\,(x\,m_{l_1} -y\,m_{l_2})+m_i\,\eta_i^+}
{2\,L^{ver}_{A^0\,h^0}}+ \frac{\eta_i^V\,(x\,m_{l_1}
-y\,m_{l_2})-m_i\, \eta_i^+}{2\,L^{ver}_{h^0\,A^0}}
\Bigg)\nonumber \\ &-& \frac{1}{2}
\,\eta_i^V\,ln\,\frac{L^{ver}_{A^0\,h^0}}{\mu^2}\,ln\,
\frac{L^{ver}_{h^0\,A^0}}{\mu^2}
\Bigg \} \nonumber \,,\\
f_3&=&-i \frac{g}{64\,\pi^2\,cos\,\theta_W} \int_0^1\,dx\,
\int_0^{1-x} \, dy \, \Bigg \{ \Bigg( (1-x-y)\,(c_V\,
\eta_i^V+c_A\,\eta_i^A)\, (x\,m_{l_1} +y\,m_{l_2}) \nonumber
\\ &+& \, m_i\,(c_A\, (x-y)\,\eta_i^-+c_V\,\eta_i^+\,(x+y))\Bigg )
\,\frac{1}{L^{ver}_{h^0}} \nonumber \\ &+& \Bigg( (1-x-y)\, (c_V\,
\eta_i^V+c_A\,\eta_i^A)\, (x\,m_{l_1} +y\,m_{l_2}) -m_i\,(c_A\,
(x-y)\,\eta_i^-+c_V\,\eta_i^+\,(x+y))\Bigg )
\,\frac{1}{L^{ver}_{A^0}} \nonumber \\ &-& (1-x-y) \Bigg
(\frac{\eta_i^A\,(x\,m_{l_1} +y\,m_{l_2})}{2}\, \Big (
\frac{1}{L^{ver}_{A^0\,h^0}}+\frac{1}{L^{ver}_{h^0\,A^0}} \Big )
+\frac{m_i\,\eta_i^-} {2} \, \Big ( \frac{1}{L^{ver}_{h^0\,A^0}}-
\frac{1}{L^{ver}_{A^0\,h^0}} \Big ) \Bigg ) \Bigg \} \,,\nonumber \\
f_4&=&-i\frac{g}{64\,\pi^2\, cos\,\theta_W} \int_0^1\,dx\,
\int_0^{1-x} \, dy \, \Bigg \{ \Bigg( (1-x-y)\,\Big ( -(c_V\,
\eta_i^A+c_A\,\eta_i^V)\, (x\,m_{l_1} -y\, m_{l_2}) \Big)
\nonumber \\ &-& m_i\, (c_A\,
(x-y)\,\eta_i^++c_V\,\eta_i^-\,(x+y))\Bigg )\,
\frac{1}{L^{ver}_{h^0}} \nonumber \\ &+& \Bigg ( (1-x-y)\,\Big (
-(c_V\, \eta_i^A+c_A\,\eta_i^V)\, (x\,m_{l_1} - y\, m_{l_2}) \Big
) + m_i\,(c_A\, (x-y)\,\eta_i^++c_V\,\eta_i^-\,(x+y)) \Bigg )
\,\frac{1}{L^{ver}_{A^0}} \nonumber \\&+& (1-x-y)\, \Bigg (
\frac{\eta_i^V}{2}\,(m_{l_1}\,x-m_{l_2}\, y)\, \, \Big (
\frac{1}{L^{ver}_{A^0\,h^0}}+\frac{1}{L^{ver}_{h^0\,A^0}} \Big )
+\frac{m_i\,\eta_i^+}{2}\, \Big (
\frac{1}{L^{ver}_{A^0\,h^0}}-\frac{1}{L^{ver}_{h^0\,A^0}} \Big )
\Bigg ) \Bigg \}\, , \label{fVAME}
\end{eqnarray}
where
\begin{eqnarray}
L^{self}_{1\,\{2\},\,h^0\,(A^0)}&=&m_{h^0\,
(A^0)}^2\,(1-x)+(m_i^2-m^2_{l_{1\,\{2\}}}\,(1-x))\,x
\nonumber \, , \\
L^{ver}_{h^0\,
(A^0)}&=&m_{h^0\,(A^0)}^2\,(1-x-y)+m_i^2\,(x+y)-k^2\,x\,y
\nonumber \, , \\
L^{ver}_{h^0\,A^0\,(A^0\,h^0)}&=&m_{A^0\,(h^0)}^2\,x+m_i^2\,(1-x-y)+
(m_{h^0\,(A^0)}^2-k^2\, x)\,y  \, ,
\label{Lh0A0}
\end{eqnarray}
and
\begin{eqnarray}
\eta_i^V&=&\xi^{E}_{N,l_1i}\xi^{E\,*}_{N,il_2}+
\xi^{E\,*}_{N,il_1} \xi^{E}_{N,l_2 i} \nonumber \, , \\
\eta_i^A&=&\xi^{E}_{N,l_1i}\xi^{E\,*}_{N,il_2}-
\xi^{E\,*}_{N,il_1} \xi^{E}_{N,l_2 i} \nonumber \, , \\
\eta_i^+&=&\xi^{E\,*}_{N,il_1}\xi^{E\,*}_{N,il_2}+
\xi^{E}_{N,l_1i} \xi^{E}_{N,l_2 i} \nonumber \, , \\
\eta_i^-&=&\xi^{E\,*}_{N,il_1}\xi^{E\,*}_{N,il_2}-
\xi^{E}_{N,l_1i} \xi^{E}_{N,l_2 i}\, . \label{etaVA}
\end{eqnarray}
The parameters $c_V$ and $c_A$ are $c_A=-\frac{1}{4}$ and
$c_V=\frac{1}{4}-sin^2\,\theta_W$. In eq. (\ref{etaVA}) the flavor
changing couplings $\xi^{E}_{N, l_ji}$ represent the effective
interaction between the internal lepton $i$, ($i=e,\mu,\tau$) and
outgoing (incoming) $j=1\,(j=2)$ one. Here we take only the $\tau$
lepton in the internal line and we neglect all the Yukawa
couplings except $\xi_{N,\tau\tau}^E$ and $\xi_{N,\tau\mu}^E$ in
the loop contributions (see Discussion section). Notice that the
parameter $\mu$ in eq. (\ref{fVAME}) is the renormalization scale,
the functions $f_1$, $f_2$, $f_3$, $f_4$ are finite and
independent of $\mu$.

The matrix element $M$ for the process $\tau\rightarrow \mu \,
\bar{\nu_i}\, \nu_i$, $i=e,\mu,\tau$ is calculated in the
framework of the model III, by connecting the $\tau\rightarrow
\mu$ transition and the $\bar{\nu_i} \nu_i$ output with the help
of the internal $Z$ boson. Finally the decay width $\Gamma$ is
obtained in the $\tau$ lepton rest frame using the well known
expression
\begin{equation}
d\Gamma=\frac{(2\, \pi)^4}{2\, m_\tau} \, |M|^2\,\delta^4
(p-\sum_{i=1}^3 p_i)\,\prod_{i=1}^3\,\frac{d^3p_i}{(2 \pi)^3 2
E_i} \,
 ,
\label{DecWidth}
\end{equation}
where $p$ ($p_i$, i=1,2,3) is four momentum vector of $\tau$
lepton ($\mu$ lepton, incoming $\nu$, outgoing $\nu$).
\section{Discussion}
In the case of vanishing charged interactions, under the
assumption that CKM type matrix in the leptonic sector does not
exist, LFV interactions arise with the help the neutral Higgs
bosons $h^0$ and $A^0$, in the framework of model III. In this
scenario the Yukawa couplings $\xi^E_{N,ij}, i,j=e, \mu, \tau$
play the main role in the determination of the $BR$ of the
processes under consideration. Since these couplings are free
parameters of the theory, they should be restricted by using the
present experimental limits of physical quantities, such as $BR$
of various leptonic decays and electric dipole moments (EDM),
anomalous magnetic moments (AMM) of leptons. In general, these
Yukawa couplings are complex and they ensure non-zero lepton EDM.

Now, we briefly discuss the case that the known light neutrinos
$\nu_i$ as massive particles and the lepton numbers $L_i$ denote
the leptons of $i^{th}$ family are not conserved. If this is so,
the lepton sector is an exact analogy to the quark sector and
there exists a similar CKM type matrix, Maki-Nakagawa-Sakata (MNS)
matrix $V_{l\nu}$ \cite{MNS}, that its elements are measured in
neutrino oscillation experiments. It has been shown that the
mixing between the muon neutrino and the heaviest mass eigenstate
of the neutrino sector, the $V_{\mu\,3}$ element, is nearly
maximal \cite{Toshito, Soudan}. The experiments on solar neutrinos
\cite{Toshito}, \cite{Ahmad} (the reactor experiments such as
CHOOZ \cite{CHOOZ}) predicted the mixing between electron neutrino
and the second heaviest mass eigenstate of the neutrino sector,
the $V_{e\,2}$ element (the heaviest mass eigenstate of the
neutrino sector, the $V_{e\,3}$ element). Notice that the corner
element $V_{e\,3}$ is much smaller than the others. On the other
hand some off-diagonal elements of MNS matrix, such as
$V_{\mu\,3}$ are large and the MNS matrix is far from diagonal in
contrast to the CKM matrix (see for example \cite{Hong} for more
discussion on lepton mixing).

With the inclusion of MNS matrix, the existence of the lepton
flavor violating (LFV) processes, $\tau\rightarrow \mu$ transition
in the present work, in the SM, would be possible. However, we
expect that the tiny masses of the internal neutrinos bring small
contribution even with the choice of maximal mixing in the
leptonic sector. (See for example \cite{Illiana}, for the
discussion of the existence of the MNS matrix and its effects on a
special LFV process). In any case, the process $\tau\rightarrow
\mu \,\bar{\nu_i}\, \nu_i$, $i=e,\mu,\tau$ should be examined
theoretically if the lepton mixing is switched on.

In the framework of the model III the $\tau\rightarrow \mu Z^*$
transition can be switched on with the internal neutral Higgs
bosons $h^0$ and $A^0$, and internal leptons $e,\mu,\tau$. This
brings unknown free parameters $\xi^{E}_{N,\mu j}$ and
$\xi^{E}_{N,\tau j},\, i,j=e,\mu,\tau$. By assuming that only the
internal $\tau$ lepton contribution is considerable, the Yukawa
couplings which does not contain $\tau$ index can be neglected.
Such a choice respects the statement that the strength of the
couplings are related with the masses of leptons denoted by the
indices of them, similar to the Cheng-Sher scenario \cite{Sher}.
Furthermore, we take $\xi^{E}_{N,ij}$ symmetric with respect to
the indices $i$ and $j$. Finally, we are left with the couplings
$\xi^{E}_{N,\tau \tau}$ and $\xi^{E}_{N,\tau \mu}$, which are
complex in general. Notice that in the following we use the
parametrization
\begin{equation}
\xi^{E}_{N,ij}= \sqrt{\frac{4 G_F}{\sqrt {2}}}
\bar{\xi}^{E}_{N,ij} \, , \label{ksipar}
\end{equation}
and we present the numerical values of some parameters used in the
calculations as a table: \\ \\

\begin{tabular}{|c|c|}
  \hline
  $m_\tau$ & 1.78 (GeV)\\
  \hline
  $m_Z$ & 91 (GeV) \\ \hline
  $m_W$ & 80 (GeV) \\ \hline
  $s_w$ & $\sqrt 0.23$ \\ \hline
  $G_F$ & $1.16637 \times 10^{-5} (GeV^{-2})\,$ \\ \hline
  $\Gamma_Z$ &  2.49  (GeV)\\ \hline
  $\Gamma_\tau$ & $2.27 \times 10^{-12}\, (GeV)$ \\
  \hline
\end{tabular}
\\ \\ \\
Table 1: The numerical values of the physical parameters used in
the numerical calculations. \\ \\

The measurement of the BRs of $\tau\rightarrow \mu \,\bar{\nu_i}\,
\nu_i$, $i=e,\mu,\tau$ decays \cite{experim} are based on counting
the number of candidate jets and correcting for efficiency and
event selection. In additions to this, the backgrounds  coming
from taus decaying to hadrons or cosmic rays should be detected.
In the process we study the output contains $\bar{\nu_i}\, \nu_i$,
$i=e,\mu,\tau$ and the extraction of this output from the most
probable one $\bar{\nu_\mu}\, \nu_\tau$ (BR $\sim 17.37\pm 0.09
\%$ \cite{PDG}), which exist in the SM theoretically, is difficult
from the experimental point of view.

In this work we studied the $BR$ of the process $\tau\rightarrow
\mu \,\bar{\nu_i}\, \nu_i$, $i=e,\mu,\tau$ and we used the
numerical value $\bar{\xi}^{E}_{N,\tau \mu}$ in the interval $5 \,
(GeV) < |\bar{\xi}^{E}_{N,\tau \mu}|<25 \, (GeV)$. Here, the upper
limit for the coupling $|\bar{\xi}^{E}_{N,\tau\mu}|$ has been
estimated in \cite{AMMmuon} as $\sim 30 \, GeV$. In this work it
is assumed that the new physics effects are of the order of the
experimental uncertainty of muon AMM, namely $10^{-9}$ by
respecting , the new experimental world average announced at BNL
\cite{BNL}
\begin{eqnarray}
a_{\mu}=11\, 659\, 203\, (8)\times 10^{-10}\,\, ,
\end{eqnarray}
which has about half of the uncertainty of previous measurements.
Here we have not used any restriction for the coupling
$\bar{\xi}^{E}_{N,\tau \tau}$ except that we choose its numerical
value larger compared to $\bar{\xi}^{E}_{N,\tau \mu}$. In addition
to this, we expect the upper limit of $\bar{\xi}^{E}_{N,\tau
\tau}$ by taking the upper limit of $\bar{\xi}^{E}_{N,\tau \mu}$
and the $BR$ of the process as $10^{-4}$-$10^{-6}$.

\begin{figure}[htb]
\vskip -3.0truein \centering \epsfxsize=6.8in
\leavevmode\epsffile{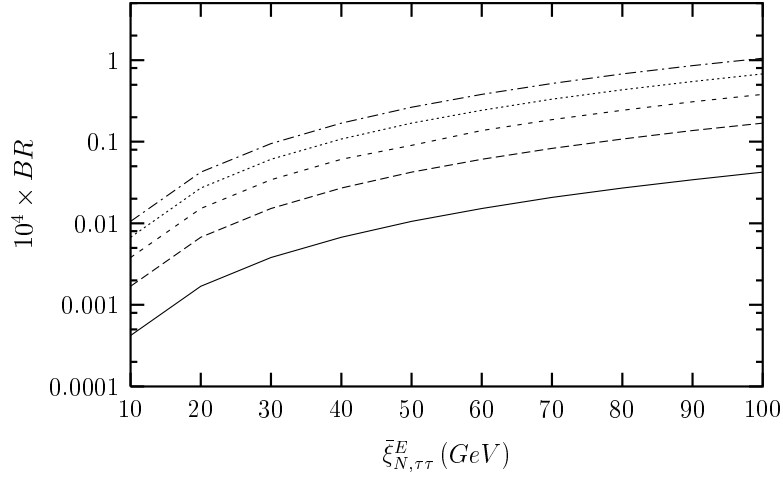} \vskip -3.0truein \caption[]{
$\bar{\xi}^{E}_{N,\tau \tau}$ dependence of the $BR$ for real
couplings and $m_{h^0}=85\, GeV$, $m_{A^0}=90\, GeV$. Here solid
(dashed, small dashed, dotted, dash-dotted) line represents the
case for $\bar{\xi}^{E}_{N,\tau \mu}=5\, GeV$ (10, 15, 20, 25
GeV). } \label{BRksitautau}
\end{figure}

\begin{figure}[htb]
\vskip -3.0truein \centering \epsfxsize=6.8in
\leavevmode\epsffile{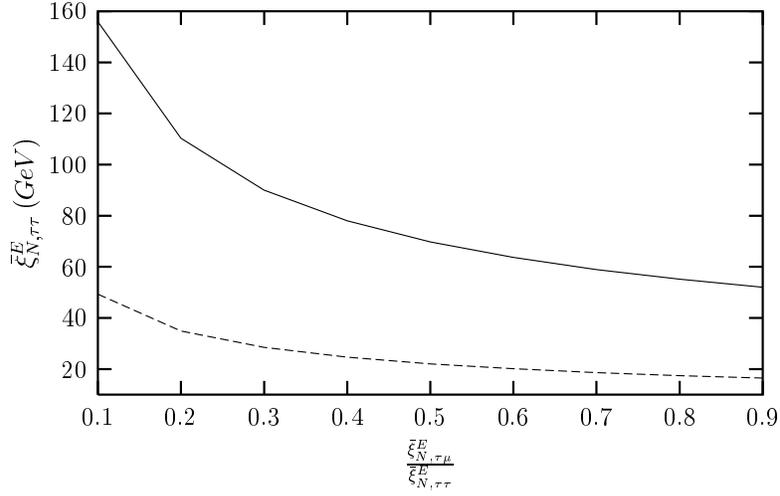} \vskip -3.0truein
\caption[]{The possible values of $\bar{\xi}^{E}_{N,\tau \tau}$
and the ratio $\frac{\bar{\xi}^{E}_{N,\tau
\mu}}{\bar{\xi}^{E}_{N,\tau \tau}}$ for $m_{h^0}=85\, GeV$,
$m_{A^0}=90\, GeV$ and the fixed values of the $BR$, $BR=10^{-4}$
(solid line) and $BR=10^{-6}$ (dashed line)}
\label{ksiDtautauMaxMin}
\end{figure}

In Fig. \ref{BRksitautau}, we present $\bar{\xi}^{E}_{N,\tau
\tau}$ dependence of the $BR$ for real couplings. Here solid
(dashed, small dashed, dotted, dash-dotted) line represents the
case for $\bar{\xi}^{E}_{N,\tau \mu}=5\, GeV$ (10, 15, 20, 25
GeV). This figure shows that the $BR$ enhances with the increasing
values of both couplings and it reaches to values at order of
magnitude of $10^{-4}$. Fig. \ref{ksiDtautauMaxMin} shows the
possible values of $\bar{\xi}^{E}_{N,\tau \tau}$ and the ratio
$\frac{\bar{\xi}^{E}_{N,\tau \mu}}{\bar{\xi}^{E}_{N,\tau \tau}}$
for the fixed values of the $BR$, $BR=10^{-4}$ (solid line) and
$BR=10^{-6}$ (dashed line). For $\frac{\bar{\xi}^{E}_{N,\tau
\mu}}{\bar{\xi}^{E}_{N,\tau \tau}}=0.1$, the $BR=10^{-4}$ is
obtained if the coupling $\bar{\xi}^{E}_{N,\tau \tau}\sim 150 \,
GeV$ and the $BR=10^{-6}$ is obtained if the coupling
$\bar{\xi}^{E}_{N,\tau \tau}\sim 50 \, GeV$. The possible
experimental search of the process $\tau\rightarrow \mu \,
\bar{\nu_i}\, \nu_i$, $i=e,\mu,\tau$ would ensure a strong clue in
the prediction of the upper limit of the coupling
$\bar{\xi}^{E}_{N,\tau \tau}$.

\begin{figure}[htb]
\vskip -3.0truein \centering \epsfxsize=6.8in
\leavevmode\epsffile{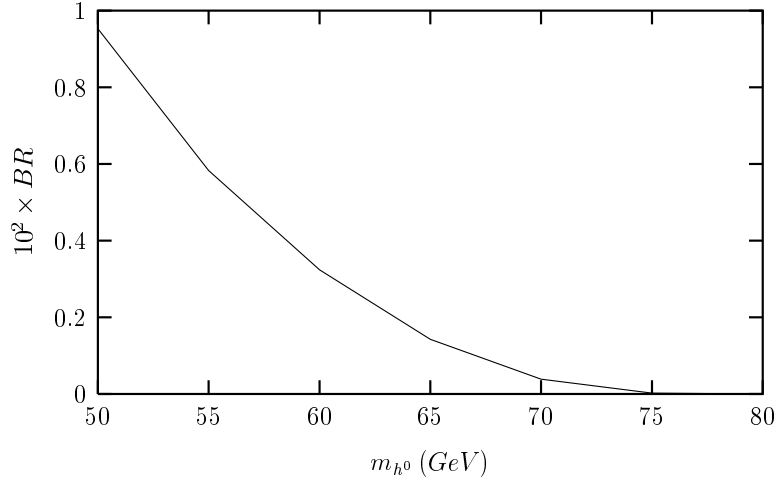} \vskip -3.0truein \caption[]{
$m_{h^0}$ dependence of the $BR$ for the fixed values of
$\bar{\xi}^{E}_{N,\tau \mu}=10\, GeV$, $\bar{\xi}^{E}_{N,\tau
\tau}=100\, GeV$ and $m_{A^0}=90\, GeV$} \label{BRmh0}
\end{figure}

Fig. \ref{BRmh0} represents the $h^0$ mass $m_{h^0}$ dependence of
the $BR$ for the fixed values of $\bar{\xi}^{E}_{N,\tau \mu}$ and
$\bar{\xi}^{E}_{N,\tau \tau}$, $\bar{\xi}^{E}_{N,\tau \mu}=10\,
GeV$, $\bar{\xi}^{E}_{N,\tau \tau}=100\, GeV$. This figure shows
that the $BR$ is sensitive to $m_{h^0}$ and it decreases with the
increasing values of $m_{h^0}$.

\begin{figure}[htb]
\vskip -3.0truein \centering \epsfxsize=6.8in
\leavevmode\epsffile{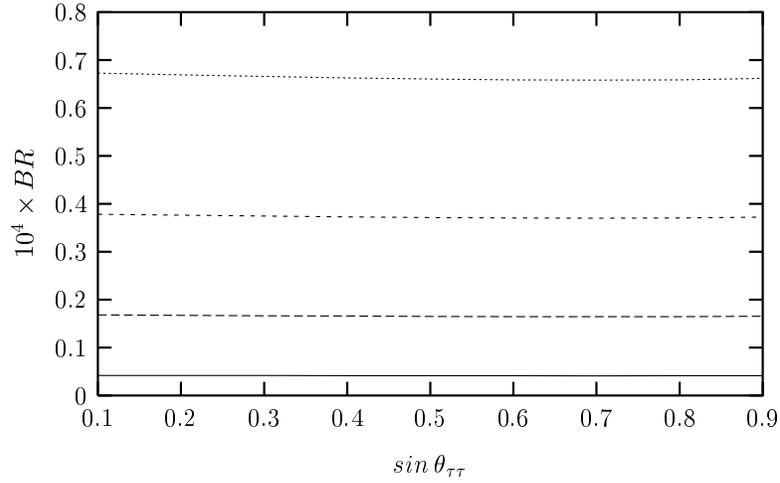} \vskip -3.0truein
\caption[]{The $\sin\, \theta_{\tau\tau}$ dependence of the $BR$
for $m_{h^0}=85\, GeV$, $m_{A^0}=90\, GeV$, $\bar{\xi}^{E}_{N,\tau
\tau}=100\, GeV$ and three different values of
$\bar{\xi}^{E}_{N,\tau \mu}$, $\bar{\xi}^{E}_{N,\tau \mu}=5,10,15,
20 GeV$ (dashed, small dashed, dotted lines)} \label{BRsintautau}
\end{figure}

Now, we take the coupling $\bar{\xi}^{E}_{N,\tau \tau}$ complex
\begin{equation}
\bar{\xi}^{E}_{N,\tau \tau}=|\bar{\xi}^{E}_{N,\tau \tau}|\, e^{i
\theta_{\tau\tau}} \, ,
\label{xicomplex}
\end{equation}
and present the $\sin\,{\theta_{\tau\tau}}$ dependence of the $BR$
for $\bar{\xi}^{E}_{N,\tau \tau}=100\, GeV$ for four different
values of $\bar{\xi}^{E}_{N,\tau \mu}$, namely
$\bar{\xi}^{E}_{N,\tau \mu}=5,10,15,20\, GeV$ (solid, dashed,
small dashed, dotted lines) in Fig. \ref{BRsintautau}. From this
figure it can be shown that the $BR$ is not sensitive to the
complexity of the coupling $\bar{\xi}^{E}_{N,\tau \tau}$.

At this stage we would like to summarize our results:

\begin{itemize}

\item We predict the $BR$ at the order of the magnitude of
$10^{-6}-10^{-4}$ for the range of the couplings,
$\bar{\xi}^{E}_{N,\tau \tau}\sim 30-100 \, GeV$ and
$\bar{\xi}^{E}_{N,\tau \mu}\sim 10-25\, GeV$. We predict the upper
limit of the coupling for the $\tau-h^0 (A^0)-\tau$ transition as
$\sim 0.3$ in the case that the $BR$ is $\sim 10^{-6}$. With the
possible experimental measurement of the process $\tau\rightarrow
\mu \, \bar{\nu_i} \,\nu_i$, $i=e,\mu,\tau$, a strong clue in the
prediction of the upper limit of the coupling
$\bar{\xi}^{E}_{N,\tau \tau}$ would be obtained. Notice that the
upper limit of the coupling $\bar{\xi}^{E}_{N,\tau \mu}$ could be
predicted previously (see for example \cite{AMMmuon}). This
analysis also ensures a hint for the physics beyond the SM.

\item We observe that the $BR$ is sensitive to the neutral Higgs
masses $m_{h^0}$ and $m_{A^0}$
\item We observe that the $BR$ is not sensitive to the possible
complexity of the Yukawa couplings.
\end{itemize}

Therefore, the future theoretical and experimental investigations
of the process $\tau\rightarrow \mu  \, \bar{\nu}_i \nu_i$ would
be informative in the determination of the physics beyond the SM
and the free parameters existing in this model.
\section{Acknowledgement}
This work has been supported by the Turkish Academy of Sciences in
the framework of the Young Scientist Award Program.
(EOI-TUBA-GEBIP/2001-1-8)

\newpage
\newpage
\end{document}